\documentclass[12pt,draftcls,onecolumn]{ieeetran}
\usepackage{slashbox}
\usepackage{amssymb}
\usepackage{amsmath}
\usepackage{graphicx}
\usepackage{slashbox}
\usepackage{array}
\usepackage{framed}
\usepackage{color}
\usepackage{graphicx}          
\usepackage{balance}
\usepackage{amsfonts}

\newtheorem{Theorem 1}{Theorem}
\newtheorem{Proposition 1}{Proposition}
\newtheorem{Theorem 2}[Theorem 1]{Theorem}
\newtheorem{Theorem 3}[Theorem 1]{Theorem}
\newtheorem{Theorem 4}[Theorem 1]{Theorem}
\newtheorem{Theorem 5}[Theorem 1]{Theorem}
\newtheorem{Theorem 6}[Theorem 1]{Theorem}
\newtheorem{Theorem 7}[Theorem 1]{Theorem}
\newtheorem{Theorem 8}[Theorem 1]{Theorem}
\newtheorem{Assumption 1}{Assumption}
\newtheorem{Assumption 2}[Assumption 1]{Assumption}
\newtheorem{Assumption 3}[Assumption 1]{Assumption}
\newtheorem{Assumption 4}[Assumption 1]{Assumption}
\newtheorem{Assumption 5}[Assumption 1]{Assumption}
\newtheorem{Remark 1}{Remark}
\newtheorem{Remark 2}[Remark 1]{Remark}
\newtheorem{Remark 3}[Remark 1]{Remark}
\newtheorem{Remark 4}[Remark 1]{Remark}
\newtheorem{Remark 5}[Remark 1]{Remark}
\newtheorem{Remark 6}[Remark 1]{Remark}
\newtheorem{Remark 7}[Remark 1]{Remark}
\newtheorem{Remark 8}[Remark 1]{Remark}
\newtheorem{Remark 9}[Remark 1]{Remark}
\newtheorem{Remark 10}[Remark 1]{Remark}
\newtheorem{Remark 11}[Remark 1]{Remark}
\newtheorem{Lemma 1}{Lemma}
\newtheorem{Lemma 2}[Lemma 1]{Lemma}
\newtheorem{Lemma 3}[Lemma 1]{Lemma}
\newtheorem{Definition 1}{Definition}
\newtheorem{Definition 2}[Definition 1]{Definition}
\newtheorem{Definition 3}[Definition 1]{Definition}
\newtheorem{Definition 4}[Definition 1]{Definition}
\newtheorem{Definition 5}[Definition 1]{Definition}
\newtheorem{Problem 1}{Problem}

\begin{document}

\title{Collective oscillation period of inter-coupled  biological negative cyclic feedback oscillators}
\author{Yongqiang Wang, {\it Senior Member, IEEE}, Yutaka Hori, {\it Member, IEEE}, Shinji Hara, {\it Fellow, IEEE}, Francis J. Doyle III, {\it Fellow, IEEE}\thanks{A special case of the work (inter-coupled Goodwin oscillators)
was published in  IEEE CDC 2012 \cite{wang_CDC:12}. The work was
supported in part by  NIH (GM096873), ICB (W911NF-09-0001,
W911NF-09-D-0001-0027) from   U.S. ARO, JSPS (23-9203), and GASR (A)
(21246067). Y. Wang, F. Doyle  are with Department of Chemical
Engineering, University of California, Santa Barbara, 93106 USA.
E-mail: wyqthu@gmail.com, frank.doyle@icb.ucsb.edu. Y. Hori and S.
Hara are with Department of Information Physics and Computing, The
University of Tokyo, Tokyo 113-8656 Japan. E-mail:
\{Yutaka\_hori,Shinji\_hara\}@ipc.i.u-tokyo.ac.jp}
} \maketitle      
\vspace{-0.5cm}
\begin{abstract}
A number of biological rhythms originate from networks comprised of
multiple  cellular oscillators. But analytical results are still
lacking on the collective oscillation period of inter-coupled gene
regulatory oscillators, which, as has been reported, may be
different from that of an autonomous   oscillator. Based on cyclic
feedback oscillators, we analyze the collective oscillation pattern
of coupled cellular oscillators. First we give a condition under
which the oscillator network exhibits oscillatory and synchronized
behavior. Then we estimate the collective oscillation period based
on a novel multivariable harmonic balance technique. Analytical
results are derived in terms of biochemical parameters, thus giving
insight into the basic mechanism of biological oscillation and
providing guidance
in synthetic biology design. 
\end{abstract}
%

\section{Introduction}

Diverse biological rhythms are generated by multiple cellular
oscillators that operate synchronously. In systems ranging from
circadian rhythms to segmentation clocks, it remains a
 challenge to understand how collective oscillation patterns
(e.g., period, amplitude) arise from
 autonomous cellular oscillations. As has been reported in the
 literature, there can be significant differences between collective oscillation patterns and
cell autonomous oscillation patterns. The differences are embodied
not only in the oscillation amplitude \cite{popovych:11}, but also
in the oscillation period  \cite{wunsche:05,Herrgen:10}.

Negative feedback is at the core of many biological oscillators
\cite{Novak:08}. One widely studied  feedback mechanism in
biological oscillators is the cyclic feedback of a sequence of
biochemical reactions, where each reaction product activates the
subsequent reaction while the end-product inhibits the  first
reaction \cite{Fall:05}. This type of structure is not only used to
formulate enzymatic control processes, but is also found in
metabolic and cellular signaling pathways \cite{Stephanopoulos:98}.
An advantage of such an oscillator is that it allows for an
analytical understanding of basic dynamical mechanisms. For example,
the oscillation conditions of a single negative cyclic feedback
oscillator were obtained in \cite{Griffith:68,
Hunding:74,Tyson:75,Hori:11}. The synchronization condition for a
network of such oscillators was reported in \cite{Hamadeh:12}. The
 oscillation patterns of a single such oscillator were also obtained in
 \cite{Rapp:76, Hori:13}. This is an important step toward
 understanding the period determination in biochemical oscillators.
 However, it remains a challenge to determine the periods
 in biological rhythms  generated by \textbf{multiple}  cellular
 oscillators. Recently, using the phenomenological phase
 model, the authors in \cite{Liu:97} proved that if intercellular coupling is weak,  the collective
 period is identical to the autonomous period. However, since the phase model contains no direct biological
mechanism of cellular clocks, its utility is limited when it comes
to checking scientific hypotheses.

 This paper analyzes the collective
 period of inter-coupled
 negative  cyclic feedback
 oscillators. 
The key idea is to decompose the whole system into scalar subsystems
and then use a multivariable harmonic
 balance technique. The multivariable harmonic
 balance technique has been adopted in  \cite{Iwasaki:08} to study central pattern generators. However, since \cite{Iwasaki:08}  assumes that the average value of oscillation
 is zero, its results are not applicable to gene regulatory oscillators. This is because, firstly, variables in gene regulatory oscillators denote concentrations of chemical reactants and cannot be negative,  thus
 do not have zero average values; secondly, as indicated in
 \cite{Rapp:76}, the zero-average-value assumption is only true when the nonlinearity is odd, which is not the case here. In this paper, we developed a new multivariable harmonic balance technique
that is applicable to gene regulatory oscillators.  Due to the
removal of the zero-average-value assumption, the  harmonic balance
equations become very difficult to solve. Here we are interested in
the collective period, so we circumvent the problem by concentrating
on synchronized oscillations. To this end, we also give an
oscillation/synchronization condition. It is worth noting that the
oscillation condition for coupled oscillators is different from that
of
 a single oscillator, as diffusive coupling may
lead to oscillations in an otherwise stable system \cite{Turing:52}.

It is worth noting that although our previous results \cite{wang:13}
gave an estimation for the collective oscillation period of a {\bf
special} type of  biological cyclic feedback oscillators connected
in a {\bf restrictive all-to-all} manner, systematic studies are
still lacking for the collective oscillation analysis of {\bf
general} cyclic feedback oscillators coupled with {\bf general}
intercellular interactions. This paper is an endeavor in this
direction. We give a method to decompose the network dynamics under
a general coupling structure, which is the key to derive the
results. This paper builds on the results in \cite{wang:13} in a
number of important ways: 1) the single oscillator model is
 more general; 2) distributed delays can be
accommodated, which is more practical \cite{MacDonald:78} than the
discrete time lag; 3) intercellular coupling is diffusive rather
than mutual repressive, and the interaction structure is more
general than the all-to-all structure in \cite{wang:13}; 4) a
synchronization condition is given, which is not discussed in
\cite{wang:13}; 5) a framework is developed to study the stability
of oscillations at the estimated frequency.
\section{Model description and decomposition}
\subsection{The model of a single oscillator}  We first consider the dynamics of a
{\bf single} negative cyclic feedback oscillator \cite{Tyson:78}:
\begin{equation}\label{eq:original
model} \left\{
\begin{aligned}
d[\mathsf{P}_1]/dT&=\rho_0/(1+[\mathsf{P}_M/K_0]^p)-k_1[\mathsf{P}_1]\\
d[\mathsf{P}_m]/dT&=\rho_{m-1}[\mathsf{P}_{m-1}]-k_m[\mathsf{P}_m],\quad m=2,3,\ldots,M\\
\end{aligned}
\right.
\end{equation}
Here $[\mathsf{P}_m]\in \mathbb{R}^1$  is the concentration of the
product $\mathsf{P}_m$ (e.g., mRNA, protein, metabolite) in the
$m$th reaction ($1\leq m\leq M$); $\rho_m$ ($0\leq m\leq M-1$) are
the rates of synthesis; $k_m$ ($1\leq m\leq M$) are degradation
rates; $1/K_0$ is the binding constant of the end product to the
transcription factor; and $p$ is the Hill coefficient, which
describes the cooperativity of end product repression.
\begin{Remark 1}
The cyclic feedback in (\ref{eq:original model}) has been used to
model the oscillations in various enzymatic control processes
\cite{Tyson:78} and metabolic control processes
\cite{Tyson:75,Tyson:79}.
\end{Remark 1}
\begin{Remark 1}
Distributed delays involved in transcription, translation, and end
product inhibition can also be incorporated in the  negative cyclic
feedback in (\ref{eq:original model}). According to the `linear
chain trick', their cumulative effects simply amount to increasing
the length of the feedback loop and the increased length is
proportional to the average magnitude of the distributed delay
\cite{MacDonald:78}.
\end{Remark 1}

The negative cyclic feedback oscillator in (\ref{eq:original model})
can be transformed into a dimension-less form
\begin{equation}\label{eq:Goodwin oscillator}
\left\{
\begin{aligned}
dx_1/dt&= f(x_M)-b_1x_1\\
dx_m/dt&=x_{m-1}-b_mx_m,\quad m=2,3,\ldots,M
\end{aligned}
\right., \quad f(x)= \frac{1}{1+x^p}
\end{equation}
by $ \varsigma=\sqrt[M]{(\prod_{i=0}^{M-1} \rho_i)/K_0},\:\:
\nu_M=\frac{1}{K_0},\:\:\nu_{j-1}=\frac{\rho_{j-1}\nu_i}{\varsigma},\:\:
x_m=\nu_m[\mathsf{P}_m],\:\: t=\varsigma T,$ and
$b_i=\frac{\rho_i}{\varsigma} $ \cite{Tyson:78}.

Transformation from (\ref{eq:Goodwin oscillator}) to
(\ref{eq:original model}) reduces parameters and thus facilitates an
analytical treatment.
\subsection{The model of interconnected oscillators}
Next we consider a network of $N$ oscillators with each oscillator
described by (\ref{eq:Goodwin oscillator}) (cf. Fig. \ref{fg:goodwin
oscillator}). Following \cite{Neda:08}, we assume that one
synchronizing factor (the $k$th
 reaction product $x_k$ ($2\leq k\leq M$)) connects the oscillators
by diffusion. Then the network dynamics is given by
\begin{equation}\label{eq:Goodwin oscillator_entrain_network}
\left\{
\begin{aligned}
dx_{1,i}/dt=&f(x_{M,i})-b_1x_{1,i}\\
dx_{m,i}/dt=&x_{m-1,i}-b_mx_{m,i},\quad 2\leq m\leq M,\:m\neq k\\
dx_{k,i}/dt=&x_{k-1,i}-b_kx_{k,i}-\sum\limits_{j=1,j\neq
i}^{N}a_{i,j}(x_{k,i}-x_{k,j})
\end{aligned}
\right.
\end{equation}
where $i=1,2,\ldots,N$ denotes the index of oscillator $i$, and
$a_{i,j}\geq 0$  denotes the coupling strength  between oscillators
$i$ and  $j$. If $a_{i,j}=0$, then there is no interaction between
oscillators $i$  and  $j$.
\begin{figure}[!hbp]
\begin{center}
  \includegraphics[width=0.50\columnwidth]{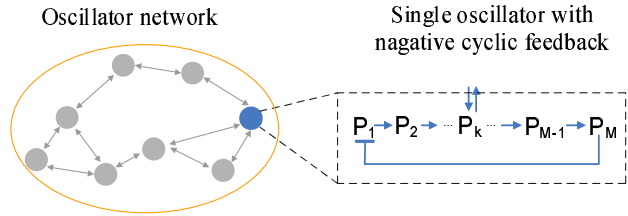}
 \vspace{-0.45cm}
    \caption{{A network of $N$ oscillators. In each oscillator,}
    $\mathsf{P}_i$ ($1\leq i\leq M-1$) activates $\mathsf{P}_{i+1}$, $\mathsf{P}_M$ inhibits the production of $\mathsf{P}_1$.}
    \label{fg:goodwin oscillator}
\end{center}
\end{figure}
 \vspace{-0.45cm}
\begin{Assumption 1}\label{as:assumption connected}
We assume $a_{i,j}=a_{j,i}$, which follows from the characteristics
of diffusion processes. We also assume connected interaction, i.e.,
there is a multi-hop path (i.e., a sequence with nonzero
$a_{i,m_1},\,a_{m_1,m_2},\,\ldots,\,a_{m_{p-1},m_{p}},\,a_{m_p,j}$)
from each node $i$ to every other node $j$.
\end{Assumption 1}
\begin{Remark 1}
Assumption \ref{as:assumption connected} is quite general. The
commonly used all-to-all interaction \cite{popovych:11}, nearest
neighbor interaction \cite{Herrgen:10}, and grid interaction
\cite{To:07} all satisfy Assumption \ref{as:assumption connected}.
\end{Remark 1}

For convenience in analysis, we can recast (\ref{eq:Goodwin
oscillator_entrain_network}) in the following matrix form:
\begin{equation}\label{eq:matrix form}
\left\{
\begin{aligned}
dX_{1}/dt=&\vec{f}(X_M)-b_1X_1\\
dX_{m}/dt=&X_{m-1}-b_mX_{m},\quad 2\leq m\leq M,\:m\neq k\\
dX_{k}/dt=&X_{k-1}-b_kX_{k}-AX_k
\end{aligned}
\right.,\qquad
X_m\hspace{-0.1cm}=\hspace{-0.1cm}\left[\hspace{-0.15cm}\begin{array}{c}x_{m,1}\\x_{m,2}\\\vdots\\x_{m,N}\end{array}\hspace{-0.15cm}\right]\in\mathbb{R}^{N\times1}
\end{equation}
\vspace{-0.1cm}
\begin{equation}\label{eq:A}
\vec{f}(X_M)\hspace{-0.1cm}=\hspace{-0.1cm}\left[\hspace{-0.15cm}\begin{array}{c}
f(x_{M,1})\\
f(x_{M,2})\\\vdots\\f(x_{M,N})\end{array}\hspace{-0.15cm}\right]\hspace{-0.2cm}\in\mathbb{R}^{N\times1},
A\hspace{-0.1cm}=\hspace{-0.1cm}\left[\hspace{-0.15cm}\begin{array}{cccc}\hspace{-0.05cm}
\sum_{j\neq 1}a_{1,j}&-a_{1,2}&\ldots&-a_{1,N}\\-a_{2,1}
&\hspace{-0.05cm}\sum_{j\neq
2}a_{2,j}&\ldots&-a_{2,N}\\\vdots&\ddots&\ddots&\vdots\\-a_{N,1}&\ldots&-a_{N,N-1}&\hspace{-0.05cm}\sum_{j\neq
N}a_{N,j}\end{array}\hspace{-0.15cm}\right]\hspace{-0.2cm}\in\mathbb{R}^{N\times
N}
\end{equation}

Since $A$ is symmetric and has zero row-sums, it can be diagonalized
by some  matrix $P$:
\begin{equation}\label{eq:diagonization of A}
A=P\Upsilon P^{-1},\quad
\Upsilon=\textrm{diag}(\upsilon_1,\:\upsilon_2,\ldots\:\upsilon_N)\in\mathbb{R}^{N\times
N}
\end{equation}
where $0=\upsilon_1<\upsilon_2\leq\ldots\leq\upsilon_N$. The
eigenvalue $0$ is associated with eigenvector
$[1\:\:1\:\,\ldots\:\:1]^T$ \cite{horn:85}. $\upsilon_2$ measures
the  connectivity of interaction. It is positive when interaction is
connected, and is greater when the interaction is stronger
\cite{horn:85}.

\subsection{Decomposition of the interconnected oscillator network model}\label{se:model diagonalization}
We are interested in the condition for oscillatory  dynamics of the
oscillator network in (\ref{eq:matrix form}), so it is necessary to
analyze its equilibrium. Next we show that (\ref{eq:matrix form})
has one unique equilibrium.

 At the equilibrium point,
we have $ dX_m^{\ast}/dt=0,\:m=\{1,2,\ldots,M\} $, which yields
\begin{equation}\label{eq:element-wisely}
\begin{aligned}
g({x_{M,i}^{\ast}})\triangleq
f(x_{M,i}^{\ast})-\prod_{m=1}^Mb_mx_{M,i}^{\ast} =\prod_{m=1,m\neq
k}b_m\sum_{j\neq i}a_{i,j}(x_{M,i}^{\ast}-x_{M,j}^{\ast})
\end{aligned}
\end{equation}

Since the interaction is bi-directional, i.e., $a_{i,j}=a_{j,i}$, it
follows
\begin{equation}\label{eq:sum}
\sum_{i=1}^Ng(x_{M,i}^{\ast})=\sum_{i=1}^N\sum_{j\neq
i}a_{i,j}(x_{M,i}^{\ast}-x_{M,j}^{\ast})=0
\end{equation}
Next, we prove that (\ref{eq:equality of g()}) holds
 by proving that both
$\max\limits_i\{g(x_{M,i}^{\ast})\}$ and
$\min\limits_i\{g(x_{M,i}^{\ast})\}$ are zero:
\begin{equation}\label{eq:equality of g()}
g(x_{M,1}^{\ast})=g(x_{M,2}^{\ast})=\ldots=g(x_{M,N}^{\ast})=0
\end{equation}

Suppose to the contrary that (\ref{eq:equality of g()}) does not
hold, then  $\max\limits_i\{g(x_{M,i}^{\ast})\}>0$ since
$\sum_{i=1}^Ng(x_{M,i}^{\ast})=0$ holds according to (\ref{eq:sum}).
Represent the index of the largest
$g(x_{M,i}^{\ast})$ among all $1\leq i\leq N$ as $q$. 
Then $x_{M,q}^{\ast}$ should be the smallest among
$x_{M,1}^{\ast},\,x_{M,2}^{\ast},\ldots,x_{M,N}^{\ast}$ because
$f(\bullet)$ and hence
 $g(\bullet)$ is a decreasing
function (cf. definition in (\ref{eq:element-wisely})). Therefore,
the rightmost hand side of  (\ref{eq:element-wisely}) should be
non-positive, and hence $g(x_{M,q}^{\ast})<0$. This contradicts the
fact that $g(x_{M,q}^{\ast})$ is the largest among
$g(x_{M,i}^{\ast})$ and is positive (due to the constraint in
(\ref{eq:sum})). Hence $\max\limits_i\{g(x_{M,i}^{\ast})\}=0$ holds.
Similarly, we can prove $\min\limits_i\{g(x_{M,i}^{\ast})\}=0$.
Therefore, we have (\ref{eq:equality of g()}), which further leads
to \vspace{-0.2cm}
\begin{equation}\label{eq:equality of f()}
f(x_{M,i}^{\ast})=B x_{M,i}^{\ast},\quad i=1,2,\ldots,N,\quad
B\triangleq\prod_{m=1}^Mb_m
\end{equation}

Since $f(x)$ is monotonic decreasing for $x\geq 0$, the solution to
(\ref{eq:equality of f()}) is unique and it satisfies
\begin{equation}\label{eq:equality}
x_{M,1}^{\ast}=x_{M,2}^{\ast}=\ldots=x_{M,N}^{\ast}=x_0>0,\quad
f(x_{0})=Bx_{0}
\end{equation}
 Therefore  the solution to
(\ref{eq:element-wisely}) is unique, thus the equilibrium point
 is unique.

\begin{figure}[!hbp]
\vspace{-0.8cm} 
\begin{minipage}[t]{0.40\linewidth}
\includegraphics[width=1.2\textwidth]{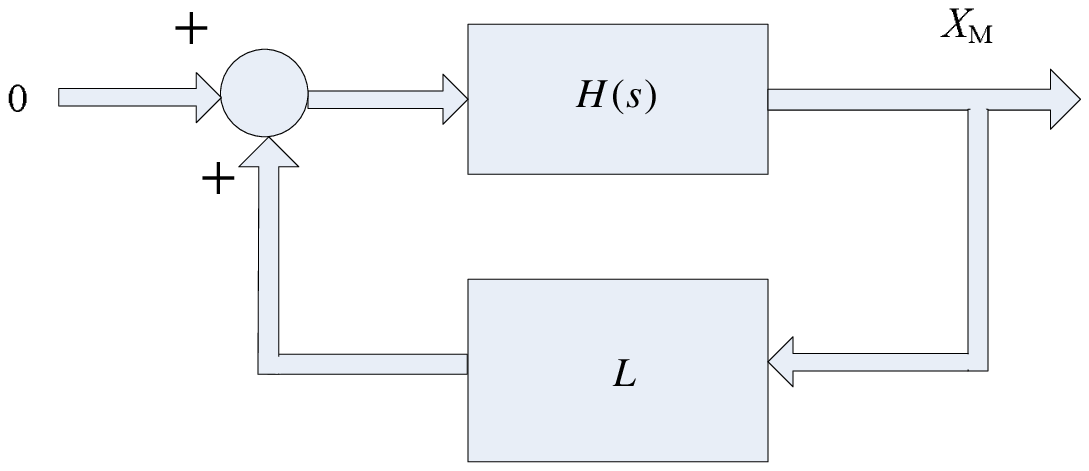}
\caption{Schematic diagram of the frequency domain formulation
($L=\sigma I$).} \label{fg:schematic}
\end{minipage}
 \hfill
\begin{minipage}[t]{0.40\linewidth}
\hspace{-1.8cm}
\includegraphics[width=1.5\textwidth]{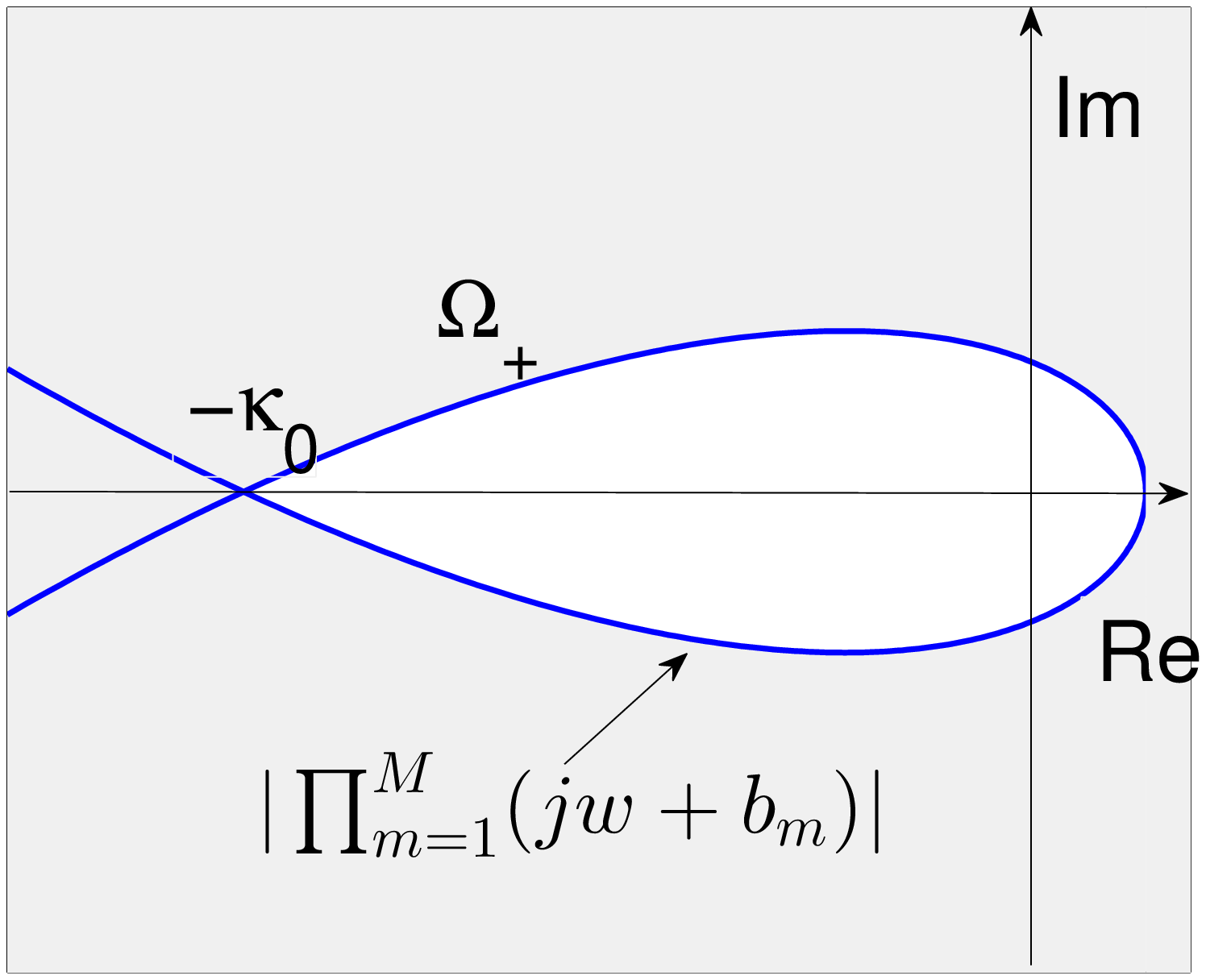}
\caption{Schematic diagram of unstable region ($\Omega_+$).
$\kappa_0=\prod_{m=1}^M\sqrt{\mu^2+b_m^2}$.} \label{fg:schematic1}
\end{minipage}
\vspace{-0.5cm}
\end{figure}

An oscillatory solution of (\ref{eq:matrix form}) needs unstable
dynamics near the equilibrium. To check the dynamics of
(\ref{eq:matrix form}) near the equilibrium, we linearize the
nonlinear item $\vec{f}(X_M)$ in (\ref{eq:matrix form}) around
$X_M^{\ast}$
\begin{equation}\label{eq:approximation}
\vec{f}(X_M-X_M^\ast)=-\sigma (X_M-X_M^\ast),\: X_M^\ast=\left[x_0\:
x_0\: \ldots\: x_0\right]^T,
\:\sigma=\frac{px_0^{p-1}}{(1+x_0^p)^2}=px_0^{p+1}B^2
\end{equation}
Then the overall dynamics of (\ref{eq:matrix form}) can be
represented in the frequency domain, as in Fig. \ref{fg:schematic},
where
\begin{eqnarray}\label{eq:H(s)_before}
H(s)&=\big((sI+b_kI+A)\prod_{m=1,m\neq k}^M(sI+b_mI)\big)^{-1}=\frac{(sI+b_kI+A)^{-1}}{\prod_{m=1,m\neq k}^{M}(s+b_m)}
\end{eqnarray}
and the matrix $L\in\mathbb{R}^{N\times N}$ denotes the influence of
the nonlinear term after linearization.

For a general matrix $L$, it is difficult to give an analytical
treatment of the dynamics in Fig. \ref{fg:schematic}. Fortunately,
under the matrix formulation in (\ref{eq:matrix form}),  we can
diagonalize the system and reduce it to multiple scalar subsystems.
This is the key to derive the analytical results in this paper.

Using (\ref{eq:matrix form}) and (\ref{eq:approximation}), we can
get $L=\sigma I\in\mathbb{R}^{N \times N}$, and hence the overall
dynamics in Fig. \ref{fg:schematic}:
\begin{equation}\label{eq:G}
G(s)=(I+\sigma H(s))^{-1}H(s) \vspace{-0.4cm}
\end{equation}
Substituting (\ref{eq:diagonization of A}) into
(\ref{eq:H(s)_before}), we have \vspace{-0.2cm}
\begin{equation}\label{eq:H(s)}
H(s)=P\Lambda
P^{-1},\:\Lambda=\textrm{diag}(\lambda_1(s),\:\lambda_2(s),\ldots\:\lambda_N(s))
\vspace{-0.2cm}
\end{equation}
 where 
$\lambda_j(s)=\frac{1}{(\prod_{m=1,m\neq k}^M (s+b_m))
(s+b_k+\upsilon_j)}$ for $j=1,2,\ldots,N.$
Substituting (\ref{eq:H(s)}) into (\ref{eq:G}) yields
\begin{equation}\label{eq:G(s)}
G(s)=P(I+\sigma\Lambda)^{-1}\Lambda P^{-1}=P\Delta P^{-1},\quad
\Delta=\textrm{diag}(\delta_1(s),\:\delta_2(s),\ldots,\:\delta_N(s))
\vspace{-0.1cm}
\end{equation}
\begin{eqnarray}\label{eq:delta_j in G(s)}
\delta_j(s)
=\frac{\lambda_j(s)}{1+\sigma\lambda_j(s)}=\frac{1}{(\prod_{m=1,m\neq
k}^M(s+b_m))(s+b_k+\upsilon_j)+\sigma},\quad j=1,2,\ldots,N
\vspace{-0.1cm}
\end{eqnarray}


So far, we have decomposed the network dynamics into multiple scalar
subsystems, which, as will be shown later, greatly facilitates an
analytical treatment of the network dynamics.

\section{Oscillation/synchronization condition}
\subsection{Theoretical analysis of the oscillation/synchronization condition}
To study the collective period, we need to guarantee that the $X_m$
in (\ref{eq:matrix form}) oscillate, and furthermore, oscillate in
synchrony. We consider the Y-oscillation, which is defined below
\cite{Iwasaki:08}:
\begin{Definition 4}
A system $\dot{x}=f(x)$ with $x(t)\in\mathcal{R}^m$ is Y-oscillatory
if each solution is bounded and there exists a state $x_i$ such that
\begin{math}\lim\limits_{\overline{_{t\hspace{-0.02cm}\rightarrow\hspace{-0.03cm}+\infty }}}x_i(t)<\overline{\lim\limits_{_{t\hspace{-0.02cm}\rightarrow\hspace{-0.05cm}+\infty}}}x_i(t)\end{math}
for almost all initial states $x(0)$.
\end{Definition 4}

To prove that (\ref{eq:matrix form}) is Y-oscillatory,  we introduce
Lemma \ref{le:lemma 1}:
\begin{Lemma 1}\label{le:lemma 1}\cite{Pogromsky:99}
System (\ref{eq:matrix form}) is Y-oscillatory if all conditions
(a), (b), and (c) hold:
\begin{itemize}
\item[(a)] It only has isolated equilibria;
\item[(b)] $\left\{X(t)\triangleq[X_1^T(t),X_2^T(t),\ldots,X_M^T(t)]^T\big|t\geq 0\right\}$ is bounded;
\item[(c)] The Jacobian matrices at
equilibria have at least one unstable eigenvalue.
\end{itemize}
\end{Lemma 1}

The result follows from these considerations: To get Y oscillations,
we need to guarantee that 1) the linearized systems near the
equilibrium points do not converge to constant values; 2) the
solutions are bounded; and, 3) there exists a homeomorphism between
solutions of the nonlinear system and its linearization. All of
these can be obtained following Theorem 1 in \cite{Pogromsky:99} and
the discussion below its proof which shows that the hyperbolicity
condition can be relaxed.

\begin{Theorem 1}\label{th:theorem 1}
The network (\ref{eq:matrix form}) has oscillatory solutions if it
satisfies the following inequality
\begin{equation}\label{eq:condition}
 R\triangleq\frac{pB(1-Bx_0)} {\kappa_0}>
 1,\quad B\triangleq\prod_{m=1}^Mb_m
\end{equation}
where $x_0$ is the  unique  positive solution to $
1/(1+x_0^p)=Bx_0$, and $\kappa_0$ is determined by
\begin{equation}\label{eq:mu}
\kappa_0=\prod_{m=1}^M\sqrt{\mu^2+b_m^2},\quad
\mu\triangleq\min\limits_{0<w<\infty} w \quad {\rm s.t.}\quad
\sum\limits_{m=1}^M \arctan(w/b_m)=\pi
\end{equation}
\end{Theorem 1}

\begin{proof}
From Lemma \ref{le:lemma 1}, the proof of Y-oscillation is
decomposed into three steps.

\subsubsection*{Step I $-$ Satisfaction of condition (a)}
As has been shown in Sec. \ref{se:model diagonalization},
(\ref{eq:matrix form}) has only one equilibrium
$X_M^{\ast}=[x_0\:x_0\:\ldots\:x_0]^T$ with $x_0>0$ determined by
$f(x_0)=B x_0$. \vspace{-0.1cm}
\subsubsection*{Step II $-$ Satisfaction
of condition (b)} (b) can be proved following the derivations in
\cite{Samad:05}.
\subsubsection*{Step III $-$ Satisfaction of condition (c)}  The Jacobian matrix having at least one eigenvalue with positive
real part is equivalent to a strictly unstable linearized system of
(\ref{eq:matrix form}) around the equilibrium,  i.e.,
(\ref{eq:G(s)}). So next we prove the strict instability of
(\ref{eq:G(s)}).
The dynamics of (\ref{eq:G(s)}) is characterized by its diagonal
elements $\delta_j(s)$ ($1\leq j\leq N$) in  (\ref{eq:delta_j in
G(s)}). For $\delta_1(s)$ we have $\upsilon_1=0$. Since both the
amplitude and argument of $\prod_{m=1}^M(s+b_m)$ increase
monotonically with the frequency on $[0,\infty)$, from graphic
analysis \cite{Hori:11} we know $\delta_1(s)$ is unstable if and
only if $-\sigma$ (defined in (\ref{eq:approximation})) is on the
left of the intersection of $\prod_{m=1}^M(jw+b_m)$ and the negative
real axis when $w$ increases from $0$ to $\infty$, i.e., $-\kappa_0$
in $(\ref{eq:mu})$ (cf. Fig. \ref{fg:schematic1}). So to have an
unstable $\delta_1(s)$ we need: \vspace{-0.2cm}
\begin{equation}\label{eq:instability_sigma}
\sigma>\kappa_0 \vspace{-0.20cm}
\end{equation}

Similarly, we have $\delta_j(s)$ ($j=2,3,\ldots,N$) is strictly
unstable if and only if $-\sigma$ is on the left of the intersection
of $(jw+b_k+\upsilon_j)\prod_{m=1,m\neq k}^M(jw+b_m)$ and the
negative real axis when $w$ increases from $0$ to $\infty$. Given
that this intersection is on the left of $-\kappa_{0}$, we know that
$G(s)$ is unstable if and only if (\ref{eq:instability_sigma})
holds. Substituting $\sigma$ in (\ref{eq:approximation}) into
(\ref{eq:instability_sigma}),
 we have
$G(s)$ is strictly unstable if and only if (\ref{eq:condition}) in
Theorem \ref{th:theorem 1} holds, i.e., condition (c) holds if
(\ref{eq:condition}) is satisfied.
\end{proof}

Next we study the condition for stable synchronized oscillations,
which is defined as:
\begin{Definition 5}\label{de:synchrony}
(\ref{eq:matrix form}) is synchronized if
$\lim\limits_{t\rightarrow+\infty}|x_{M,i}(t)-x_{M,j}(t)|=0$ holds
for any $1\leq i,j\leq N$.
\end{Definition 5}
\begin{Remark 1}
Only $x_{M,i}$ is used in the definition of synchronization. This is
because according to the modeling assumption, $x_{M,i}$ corresponds
to the concentration of inhibitor or enzyme, which can be regarded
as the output of an
 oscillator. Under this
definition, when the system is synchronized, $x_{m,i}$ ($1\leq i
\leq N$) may be identical or non-identical for $m\neq M$.
\end{Remark 1}
\begin{Theorem 1}\label{th:sync condition}
The oscillator network (\ref{eq:matrix form}) has stable
synchronized oscillations only
 if
 \vspace{-0.2cm}
 \begin{equation}\label{eq:synchornization condition}
Nz_0/(N-1)<\sqrt{\mu_2^2+(b_k+\upsilon_2)^2}\prod_{m=1,m\neq
k}^{M}\sqrt{\mu^2+b_m^2},\quad
z_0=\max\limits_{x>0}\frac{px^{p-1}}{(1+x^p)^2} \vspace{-0.15cm}
\end{equation}
is satisfied, where $\upsilon_2$ is the second smallest eigenvalue
of $A$ and  $\mu_2$ is the minimal positive solution to $
  \arctan(\mu_2/(b_k+\upsilon_2))+\sum_{m=1,m\neq
  k}^M\arctan\frac{\mu_2}{b_m}=\pi.$
\end{Theorem 1}
\begin{proof}
A necessary condition for the synchronization in Definition
\ref{de:synchrony} is the stability of synchronization manifold
$x_{M,1}(t)=x_{M,2}(t)=\ldots=x_{M,N}(t)$. So we check
$y_{m,i}\triangleq x_{m,i}-\frac{\sum_{j=1}^N x_{m,j}}{N}$, which
measures the deviation of the $i$th oscillator from the
synchronization manifold. If the dynamics of $y_{m,i}$ is stable for
all $1\leq i \leq N$, then the synchronization manifold is stable.

From the definition of $y_{m,i}$ and (\ref{eq:matrix form}), we can
get the dynamics of $y_{m,i}$:
\begin{equation}\label{eq:error_dynamics}
\left\{
\begin{aligned}
dy_{1,i}/dt=&\hbar_i-b_1y_{1,i},\quad \hbar_i=f(x_{M,i})-\sum_{j=1}^Nf(x_{M,j})/N\\
dy_{m,i}/dt=&y_{m-1,i}-b_my_{m,i},\quad 2\leq m\leq M,\:m\neq k\\
dy_{k,i}/dt=&y_{k-1,i}-b_ky_{k,i}-\sum\limits_{j=1,j\neq i}^{N}a_{i,j}(y_{k,i}-y_{k,j})\\
\end{aligned}
\right.
\end{equation}
\vspace{-0.4cm}

Linearizing the system along the synchronization manifold yields:
\vspace{-0.2cm}
\begin{equation}\label{eq:matrix form of error
dynamics}
 \left\{
\begin{aligned}
dY_{1}/dt=&KY_M-b_1Y_1\\
dY_{m}/dt=&Y_{m-1}-b_mY_{m},\quad 2\leq m\leq M,\:m\neq k\\
dY_{k}/dt=&Y_{k-1}-b_kY_{k}-AY_k\
\end{aligned}
\right., \qquad
Y_m\hspace{-0.1cm}=\hspace{-0.1cm}\left[\hspace{-0.15cm}\begin{array}{c}y_{m,1}\\y_{m,2}\\\vdots\\y_{m,N}\end{array}\hspace{-0.15cm}\right]\in\mathbb{R}^{N\times
1} \vspace{-0.1cm}
\end{equation}
In (\ref{eq:matrix form of error dynamics}), matrix $A$ is given in
(\ref{eq:A}) and $K$ is a matrix with diagonal elements given by
$-z$ and off-diagonal elements given by $\frac{z}{(N-1)}$ where
$z=\frac{px^{p-1}}{(1+x^p)^2}$.


Eqn (\ref{eq:matrix form of error dynamics}) can be described in the
frequency domain as shown in Fig. \ref{fg:schematic}, where $H(s)$
is the same as (\ref{eq:H(s)_before}) but $L$ is replaced by $L=K$.
The transfer function of (\ref{eq:matrix form of error dynamics}) is
$Q(s)=(I-H(s)K)^{-1}H(s)$. It can be verified that $A$ and $K$
commute, so we can  diagonalize them simultaneously \cite{horn:85}
and, thus diagonalize $Q(s)$ as
$Q(s)=P\textrm{diag}(q_1(s),q_2(s),\dots,q_M(s))P^{-1}$ with
\begin{eqnarray}
q_i(s)&=&\frac{1}{(s+b_k+\upsilon_i)\prod_{m=1,m\neq
k}^M(s+b_m)+\chi_i},\quad i= 1,2,\ldots,N
\end{eqnarray}
where $\chi_1=0$ and
$\chi_2=\chi_3=\ldots=\chi_N=\frac{-N}{(N-1)}z=\frac{-N}{(N-1)}\frac{px^{p-1}}{(1+x^p)^2}$
are the eigenvalues of $K$. Note that they are different at
different positions on the synchronization manifold.

Note $v_0=0$, $q_1(s)$ is stable, so we only consider $q_i(s)$ for
$i=2,3,\ldots,N$. Following Theorem \ref{th:theorem 1}, we know that
$q_i(s)$ is stable if and only if $\chi_i$ resides on the right hand
side of the intersection (denote it as $-\kappa_i$) of $q_i(s)$ with
the negative real axis, which is determined by
\begin{equation}
\kappa_i=\sqrt{\mu_i^2+(b_k+\upsilon_i)^2}\prod_{m=1,m\neq
k}^{M}\sqrt{\mu_m^2+b_m^2}
\end{equation}
where $\mu_i$ is the minimal positive solution to
$\arctan\frac{\mu_i}{b_k+\upsilon_i}+\sum_{m=1,m\neq
  k}^M\arctan\frac{\mu_i}{b_m}=\pi.$ It can be verified that $\kappa_i$ increases with $\upsilon_i$. So
if  $\chi_i>-\kappa_i$ holds for $i=2$, which corresponds to the
smallest $\upsilon_i$ among $i=2,3, \ldots,N$, then the
synchronization manifold is stable. Given that $\chi_i$ is a
function of $x$, (\ref{eq:synchornization condition}) can be
obtained by setting $\chi_2$ to its minimal value among all $x>0$.
\end{proof}
\begin{Remark 1}
Compared with the sufficient condition in \cite{Hamadeh:12}, Theorem
\ref{th:sync condition} is a necessary condition for global
synchronization. In Sec. V we use simulations to estimate its
conservativeness.
\end{Remark 1}
\vspace{-0.3cm}
\subsection{Biological insight}
It can be verified that for $M\geq 2$, $R$ in (\ref{eq:condition})
increases
 with $M$, the length of the cyclic feedback.
Given that a larger $R$ makes (\ref{eq:condition}) easier to
satisfy, a longer cyclic feedback loop (i.e., a larger $M$, meaning
involving more serial reactions) makes oscillation easier. Moreover,
recalling the positive correlation between the averaged value of
distributed delay and the length of feedback loop (cf. Remark 2), we
can  infer that a larger delay also makes oscillation easier.

From (\ref{eq:synchornization condition}), we can see that with an
increase in $z_0$, a larger $\upsilon_2$ (i.e., a stronger
intercellular interaction) is required to achieve synchronization.
Given that for $p>1$, $z_0$ can be verified an increasing function
of the Hill coefficient $p$, we know that a system having a higher
Hill coefficient (i.e., a higher cooperativity of end product
repression) requires stronger coupling to maintain synchronization.
Furthermore, we can also verify that a longer feedback chain makes
the right hand side of the inequality in (\ref{eq:synchornization
condition}) lower, and thus makes (\ref{eq:synchornization
condition}) harder to satisfy. Given the positive correlation
between the feedback loop length  and the averaged distributed delay
(cf. Remark 2), we can infer that a larger delay makes
synchronization more difficult to maintain.

\section{Oscillation period estimation based on multivariable harmonic
balance}
\subsection{Oscillation analysis based on  harmonic
balance technique}

We reformulate the problem of oscillation analysis using a
multivariable harmonic balance technique. This is motivated by the
observation that $H(s)$ is a low pass filter thus higher order
harmonics of oscillations in the closed-loop system can be safely
neglected. Hence $x_{M,i}$ can be approximated by its zero-order and
first-order harmonic components \cite{Iwasaki:08,Khalil:02}:
\vspace{-0.cm}
\begin{equation}
x_{M,i}=\alpha_i+\beta_i\sin(wt+\phi_i),\: i=1,2,\ldots,N
\end{equation}
\vspace{-0.1cm} where $\alpha_i$ and $\beta_i$ denote the amplitudes
of the zero-order and the first-order harmonic components,
respectively, and $w$ and $\phi_i$ denote the oscillation frequency
and phase, respectively.

Since $f(\bullet)$ is a static nonlinear function, it can be
approximated by describing functions \cite{Khalil:02}:
\begin{equation}
f(x_{M,i})\approx \xi_i \alpha_i +\eta_i \beta_i\sin(wt+\phi_i)
\end{equation}
\begin{equation}\label{eq:describing_eta_i}
\xi_i=\frac{1}{2\pi \alpha_i}\int_{-\pi}^{\pi}f(\alpha_i +
\beta_i\sin(t))dt,\quad \eta_i=\frac{1}{\pi
\beta_i}\int_{-\pi}^{\pi}f(\alpha_i + \beta_i\sin(t))\sin(t)dt
\end{equation}
The describing function $\xi_i$ is the gain of $f(\bullet)$ when the
input is a constant $\alpha_i$ and the output is approximated by the
zero-order harmonic component. The describing function $\eta_i$ is
the gain of $f(\bullet)$ when the input is a sinusoid of amplitude
$\beta_i$ and the output is approximated by the first-order harmonic
component \cite{Khalil:02}.

Consequently,  $\alpha_i$ and $\beta_i$ are expected to satisfy
\cite{Iwasaki:08}:
\begin{equation}\label{eq:zero-order harmonic equation}
(I- H(0)\Xi)\vec\alpha=0,\quad (I- H(jw)\Pi)\vec\beta=0
\end{equation}
where
$\Xi=\textrm{diag}\{\xi_1,\,\ldots,\xi_N\}\in\mathbb{R}^{N\times
N}$,
$\Pi=\textrm{diag}\{\eta_1,\,\ldots,\eta_N\}\in\mathbb{R}^{N\times
N}$, and
\[
\vec\alpha=\left[\begin{array}{ccccc}\alpha_1& \alpha_2&\ldots&
\alpha_N\end{array}\right]\in\mathbb{R}^{N\times 1},\:
\vec\beta=\left[\begin{array}{cccc}\beta_1 e^{j\phi_i}&
\beta_2e^{j\phi_2}&\ldots&\beta_Ne^{j\phi_N}\end{array}\right]\in\mathbb{R}^{N\times
1}.
\]
Note that
 (\ref{eq:zero-order harmonic equation}) are referred to as harmonic
balance equations.

Let $\Xi^{\ast}$ and $\Pi^{\ast}$ be matrices satisfying
(\ref{eq:zero-order harmonic equation}). Define two linear systems
$G_0(s)$ and $G_1(s)$ as
\begin{equation}\label{eq:G_0 and G_1}
\begin{aligned}
G_{0}(s) \triangleq (I-H(s)\Xi^{\ast})^{-1}H(s),\quad G_{1}(s)
\triangleq (I-H(s)\Pi^{\ast})^{-1}H(s)
\end{aligned}
\end{equation}
$G_0(s)$ and $G_1(s)$ are obtained by replacing the nonlinearity
$f(\bullet)$ with the constant gain computed from the describing
functions.  To ensure that the predicted oscillation frequency is
biologically significant, oscillations at the estimated frequency
must be stable, which, according to \cite{Glad:00}, can be ensured
if both $G_0(s)$ and $G_1(s)$ are marginally stable (the only
unstable poles are $s=0$ and $s=\pm jw$, respectively). Therefore,
estimating the collective period can be reduced to:
\begin{Problem 1}
For the given $H(s)$ and nonlinear function $f(x)$ in
(\ref{eq:Goodwin oscillator_entrain_network}), find $w$ such that
\textcircled{1} equation (\ref{eq:zero-order harmonic equation}) is
satisfied, and \textcircled{2}
 $G_0(s)$ and $G_1(s)$ in (\ref{eq:G_0 and G_1}) are marginally
stable. \vspace{-0.3cm}
\end{Problem 1}
\subsection{Oscillation period of coupled negative cyclic feedback oscillators}
Eqn (\ref{eq:zero-order harmonic equation}) is very difficult to
solve since in general $\Xi$ and $\Pi$ depend on $\vec\alpha$ and
$\vec\beta$. Keeping in mind that we are interested in the
collective period,  we concentrate on solutions that describe
synchronized oscillations. According to Definition
\ref{de:synchrony}, synchrony means that $x_{M,i}$ are identical,
i.e., 1) the phases $\phi_i$ are identical; 2) the amplitudes
$\alpha_i$ and $\beta_i$ are respectively identical. Given that
$\xi_i$ and $\eta_i$ are determined by $\alpha_i$ and $\beta_i$, we
further have the equality of all $\xi_i$ and all $\eta_i$:
\begin{equation}\label{eq:scalarization}
\vspace{-0.3cm} {\vec\alpha}=\alpha\vec{1},\:\:
{\vec\beta}=\beta\vec{1},\:\: \Xi=\xi I,\:\:\Pi=\eta I,\:
\vec{1}\triangleq\left[1\:\:1\:\:\ldots\:\:1\right]^T
\in\mathbb{R}^{N\times 1},\:
I=\textrm{diag}\{1,\,\ldots,1\}\in\mathbb{R}^{N\times N}
\end{equation}
where $\alpha$, $\beta$, $\xi$, and $\eta$ are constants. 
Hence (\ref{eq:zero-order harmonic equation}) reduces to
\begin{eqnarray}\label{eq:zero-order}
\vspace{-0.3cm}
 (\frac{1}{\xi}I-  H(0))\alpha\vec{1}=0, \quad
(\frac{1}{\eta}I- H(jw))\beta\vec{1}=0
\vspace{-0.3cm}
\end{eqnarray}
which further means that $\frac{1}{\xi}$ and $\frac{1}{\eta}$ are
the respective eigenvalues of $H(0)$ and $H(jw)$ corresponding to
the eigenvector with identical elements.

From (\ref{eq:H(s)}), we know the eigenvalues of $H(0)$ are
$\lambda_j =\frac{1}{(b_k+\upsilon_j)(\prod_{m=1,m\neq k}^Mb_m)}$
for $j=1,2,\ldots,N$. Since only $\lambda_1$ corresponds to
eigenvectors with identical elements, we have (note $\upsilon_1=0$)
 \vspace{-0.1cm}
\begin{eqnarray}\label{eq:xi}
\xi=1/\lambda_1=\prod_{m=1}^Mb_m
 \vspace{-0.1cm}
\end{eqnarray}
Similarly, we can get that the eigenvalues of $H(jw)$ are
 $\lambda_j(jw)
=\frac{1}{(jw+b_k+\upsilon_j)\prod_{m=1,m\neq k}^M(jw+b_m)}$  for
$j=1,2,\ldots,N$. Since only $\lambda_1$ corresponds to eigenvectors
with identical elements, we have
\begin{equation}\label{eq:eta_eigenvalue}
\eta=1/\lambda_1(jw)=\prod_{m=1}^M(jw+b_m)
 \vspace{-0.1cm}
\end{equation}
According to (\ref{eq:describing_eta_i}), $\eta$ is real, thus the
right hand side of (\ref{eq:eta_eigenvalue}) must be real. Given
that $\mu$ is the minimal frequency that makes
$\prod_{m=1}^M(jw+b_m)$ have zero imaginary part (angular $\pi$),
the collective frequency is determined by $\mu$ in (\ref{eq:mu}) and
\vspace{-0.1cm}
\begin{equation}\label{eq:eta}
 \eta=\prod_{m=1}^M(j\mu+b_m)=-\prod_{m=1}^M\sqrt{\mu^2+b_m^2}
 \vspace{-0.1cm}
\end{equation}

To solve Problem 1, it remains to prove that $G_0(s)$ and $G_1(s)$
in (\ref{eq:G_0 and G_1}) are marginally stable \cite{Glad:00}, or
in other words: (1) $G_0(s)$ has one pole of $s=0$ and the rest in
the open left half plane and, (2) $G_1(s)$ has imaginary poles
$s=\pm jw$ and the rest in the open left half plane.

Substituting $\Xi$ and $\Pi$ in (\ref{eq:scalarization}) into
(\ref{eq:G_0 and G_1}) yields
\begin{equation}\label{eq:G_0}
G_0(s)=(I-\xi H(s))^{-1}H(s) ,\quad  G_1(s)=(I-\eta H(s))^{-1}H(s)
\end{equation}
with $\xi$ and $\eta$ given in (\ref{eq:xi}) and (\ref{eq:eta}),
respectively.

First consider $G_0(s)$. From (\ref{eq:G(s)})-(\ref{eq:delta_j in
G(s)}), we know that the eigenvalues of $G_0(s)$ are given by
\begin{equation}\label{eq:eigenvalues}
\begin{aligned}
\delta_j(s) =\frac{1}{(s+b_k+\upsilon_j)\prod_{m=1,m\neq
k}^M(s+b_m)-\xi},\quad j=1,2,\ldots,N
\end{aligned}
\end{equation}
 Substituting $\xi$ in (\ref{eq:xi}) into (\ref{eq:eigenvalues}), we know that the poles of $G_0(s)$ in (\ref{eq:G_0}) are
 the roots of
\begin{equation}\label{eq:roots G_0(2)}
 (s+b_k+\upsilon_j)\prod_{m=1,m\neq k}^M(s+b_m)-\prod_{m=1}^Mb_m=0,\quad j=1,2,\ldots,N
\end{equation}
For $j=1$, since $\upsilon_1=0$,  (\ref{eq:roots G_0(2)}) has one
root $s=0$. It can also be derived that all the rest of the roots
have negative real parts since for all $s$ with a positive real
part, the modulus of $\prod_{m=1}^Mb_m$ is less than
$\prod_{m=1}^M(s+b_m)$, which
 makes equality in (\ref{eq:roots G_0(2)}) impossible. Similarly, we can get that for $j\neq 1$, all roots of (\ref{eq:roots G_0(2)})
have negative real parts. Hence $G_0(s)$ is marginally stable.

Following the same line of reasoning, we can prove that the
eigenvalues of $G_1(s)$ are
\[
\delta_j(s)
=\frac{1}{(s+b_k+\upsilon_j)\prod_{m=1}^M(s+b_m)-\eta},\quad
j=1,2,\ldots,N
\]
 with $\eta$ given in (\ref{eq:eta}). And hence
 its poles are determined by the roots of
\begin{equation}\label{eq:roots G_1(2)}
\begin{aligned}
 (s+b_k+\upsilon_j)\prod_{m=1,m\neq k}(s+b_m)+\prod_{m=1}^M\sqrt{\mu^2+b_m^2}=0,\quad j=1,2,\ldots,N
\end{aligned}
\end{equation}
where $\mu$ is given in (\ref{eq:mu}).

For $j=1$, we can verify that  $s=\pm jw$ are roots of
(\ref{eq:roots G_1(2)}). We can also verify that all the other roots
of (\ref{eq:roots G_1(2)}) are stable,  since for all $s$ with a
positive real part, the intersection of $\prod_{m=1}^M(s+b_m)$ and
the negative real axis is less than
$-\kappa_0\hspace{-0.1cm}=\hspace{-0.1cm}-\hspace{-0.1cm}\prod\limits_{m=1}^M\hspace{-0.15cm}\sqrt{\mu^2+b_m^2}$,
which makes the equality in (\ref{eq:roots G_1(2)}) impossible.
Similarly, for $j\neq 1$, we can derive that all roots of
(\ref{eq:roots G_1(2)}) are in the open left half plane. So $G_1(s)$
is marginally stable. Hence oscillations at the derived frequency
$\mu$ are stable.

\begin{Proposition 1}\label{th:proposition}
The solution for the oscillation frequency $w$ in Problem 1 is given
by $w=\mu$ where $\mu$ is defined in (\ref{eq:mu}).
\end{Proposition 1}

From the above derivation, we can see that the collective
oscillation period is expected to be
\begin{equation}\label{eq:collective frequency}
T_{\rm{collective}}=2\pi/\mu \vspace{-0.35cm}
\end{equation}

\subsection{Biological insight}
The collective period in (\ref{eq:collective frequency}) is given in
terms of the dimensionless parameters in (\ref{eq:Goodwin
oscillator}). The actual collective frequency in dimensional
parameters are given by $\Omega=\varsigma\mu$ where $\mu$ is the
minimal positive solution to $ \sum\limits_{m=1}^M
\arctan\frac{\mu}{b_m}=\sum\limits_{m=1}^M
\arctan\frac{\mu}{k_m/\varsigma}=\pi. $
 So the actual collective
frequency is the minimal positive solution to $\sum\limits_{m=1}^M
\arctan\frac{\Omega}{k_m}=\pi$. This means that the collective
frequency $\Omega$ increases with an increase in the degradation
rate of each component ($k_m$), but it is independent of the rates
of transcription, translation, and synthesis. These give insights
into the basic determination mechanism of the collective period  in
coupled biological oscillators, and may further provide guidance in
synthetic biology design.

From the above derivation, we can see that under interaction
(\ref{eq:Goodwin oscillator_entrain_network}), the collective period
is only determined by $k_m$ ($m=1,2,\ldots,M$), and it is
independent of intercellular coupling. The results are obtained
based on analytical treatment of a network of coupled gene
regulatory oscillators and they corroborate the results in
\cite{Liu:97}, which are obtained using the phenomenological
single-variable phase model and state that the strength of
intercellular coupling does not affect the collective period of
circadian rhythm oscillator networks. In fact, this is reasonable
since the coupling is similar to the linear consensus protocol
\cite{Olfati-sater:07}, which only affects the process to
synchronization. Moreover, when the degradation rate is fixed, it
can be inferred that  $\Omega$ decreases with an increase in the
length of the feedback loop $M$. Therefore, a longer feedback loop
corresponds to a longer collective period. Furthermore, recall that
the effect of distributed delay amounts to increasing the length of
the feedback loop and the increased length is proportional to the
averaged delay, hence, a larger delay in individual loops means a
longer collective period.

\begin{Remark 4}
If the coupling is different from (\ref{eq:Goodwin
oscillator_entrain_network}), it may affect the collective period,
as exemplified by the mutual repressive coupling in  \cite{wang:13}.
\end{Remark 4}

\section{Numerical study}
 We considered a network of nine oscillators coupled via the second reactant. The coupling strengths $a_{i,j}$ were
 chosen from a uniform distribution on $[0,\,20]$ and the coupling
topology is verified to be connected.
 First we tested our oscillation condition,  
with  results given in Table I. It can be seen that oscillation can
be obtained only when the parameters satisfy $R> 1$ in
(\ref{eq:condition}).
\begin{table}[htb]
\centering
\begin{minipage}[t]{\linewidth}
\centering \vspace{-0.25cm}
 \label{ta:table1} \caption{Test of the  oscillation condition}
\vspace{-0.45cm}
\begin{tabular}{||c|c|c|c|c|c|c|c|c|c|c|c||}
 \hline\hline
$p$&$b_1$&$b_2$&$b_3$&$b_4$&$b_5$&$b_6$&$b_7$&$b_8$&$b_9$&$R$&Simulation
results
\\
\hline 3 & 0.5 &0.5 & 0.5 & 0.5 &0.5 & 0.5& 0.5 &0.5 & 0.5 &1.6898 & Oscillation \\
\hline 3 & 0.5 &0.6 & 0.7 & 0.8 &0.9 & 0.8& 0.7 &0.6 & 0.5 &1.5733 & Oscillation\\
\hline 3 & 0.7 &0.7 & 0.7 & 0.7 &0.7 & 0.7& 0.7 &0.7 & 0.7 &1.5571 & Oscillation\\
\hline 3 & 0.8 &0.8 & 0.8 & 0.8 &0.8 & 0.8& 0.8 &0.8 & 0.8 &1.3549 & Oscillation\\
\hline 3 & 0.88 &0.88 & 0.88 & 0.88 &0.88 & 0.88& 0.88 &0.88 & 0.88 &1.0707 & Oscillation \\
\hline 3 & 0.9 &0.9 & 0.9 & 0.9 &0.9 & 0.9& 0.9 &0.9 & 0.9 &0.9819 & No oscillation \\
\hline 3 & 1.0 &1.0 & 1.0 & 1.0 &1.0 & 1.0& 1.0 &1.0 & 1.0 &0.4721 & No oscillation \\
\hline\hline
\end{tabular}
\vspace{-0.5cm}
\end{minipage}
\end{table}

We then compared our synchronization condition with the sufficient
synchronization condition in \cite{Hamadeh:12}. The gap between the
two conditions is shown in Table II. It is worth noting that
 extensive numerical simulations showed that our synchronization condition is
 minimally conservative because despite the fact that it is a necessary synchronization condition, it successfully ensured synchronization
for all $10^6$ runs with initial conditions randomly chosen from the
interval $[0,\,10^3]$.

\begin{table}[htb]
\centering
\begin{minipage}[t]{\linewidth}
\centering
 \label{ta:table1} \caption{Comparison of the required network connectivity $\upsilon_2$ to achieve synchronization}
\vspace{-0.35cm}
\begin{tabular}{||c|c|c|c|c|c|c|c|c|c||}
\hline\hline $b_1=b_2=\ldots=b_9$&$0.50$&0.55&$0.60$&0.65&$0.70$&0.75&$0.80$&0.85\\
\hline The required $\upsilon_2$ in \cite{Hamadeh:12}&  178.13 &82.78& 40.94 &21.24& 11.40 &6.22& 3.35&1.71 \\
\hline The required $\upsilon_2$ in this paper             & 127.98 &59.45& 29.36 & 15.19 &8.11 &4.37& 2.30&1.09\\
\hline\hline
\end{tabular}
\vspace{-0.5cm}
\end{minipage}
\end{table}

We also verified the estimated collective periods in oscillatory
cases. The results (in Table III) show that the estimated values
approximate the actual collective periods closely.
\begin{table}[htb]
\centering
\begin{minipage}[t]{\linewidth}
\centering
 \label{ta:table1} \caption{Comparison between the estimated collective period [s] and the actual collective period [s]}
\vspace{-0.35cm}
\begin{tabular}{||c|c|c|c|c|c|c|c|c|c||}
\hline\hline $b_1=b_2=\ldots=b_9$&$0.50$&0.55&$0.60$&0.65&$0.70$&0.75&$0.80$&0.85\\
\hline Actual value &  40.9 &36.2& 32.3 &29.0& 26.2 &23.9& 22.03&20.4 \\
\hline Estimated value  & 36.0 &32.7& 30.0 & 27.7&25.7 &24.0& 22.5&21.1 \\
\hline Estimation error & -11.9\%&-9.67\% & -7.12\%&-5.86\% & -1.91\%&0.42\%&0.89\%&3.4\% \\
\hline\hline
\end{tabular}
\vspace{-0.5cm}
\end{minipage}
\end{table}

\section{Conclusions}
Biological rhythms are generated by networks of interacting cellular
oscillators. The mechanisms that describe how the collective
oscillation patterns arise from autonomous cellular oscillations are
poorly understood. Based on a network of coupled negative cyclic
feedback oscillators, we studied the oscillation/synchronization
condition and collective period of coupled biochemical oscillators
by using a multivariable harmonic balance technique. We gave
oscillation and synchronization conditions of coupled negative
cyclic feedback oscillators. We also analytically estimated the
collective oscillation period of the oscillator network and examined
how it is affected by the parameters of biochemical reactions. The
results are confirmed by numerical simulations and can provide
guidance in synthetic oscillator design in biology.

\bibliographystyle{unsrt}
\bibliography{abbr_bibli}

\end{document}